# Efficeint FPGA Design of 32-bit Euclid's GCD based on Sum of Absolute Difference

Saeideh Nabipour
Department Computer and Electrical Engineering, Ahrar Institute of Technology and Higher Education
Rasht, IRAN
Saeideh.nabipour@gmail.com

Masoume Gholizade
Department Computer and Electrical Engineering, University of Semnan
Semnan, IRAN
masoume_gholizade@semnan.ac.ir

Nima Nabipour
Department Computer and Electrical Engineering, University of Azad Rasht
Rasht, IRAN
Nima.nabipour@gmail.com

*Abstract*—Euclid's algorithm is widely used in calculating of GCD (Greatest Common Divisor) of two positive numbers. There are various fields where this division is used such as channel coding, cryptography, and error correction codes. This makes the GCD a fundamental algorithm in number theory, so a number of methods have been discovered to efficiently compute it. The main contribution of this paper is to propose a new method named Optimized-GCDSAD that computes the GCD of two 32-bit numbers by utilizing Sum of Absolute Difference (SAD) block which is based on a fast carry-out generation function. The efficiency of the proposed architecture is evaluated based on criteria such as time (latency), area delay product (ADP) and space (slice number) complexity. The VHDL codes of these architectures have been implemented and synthesized through ISE 14.7 on five different Xilinx chips The results demonstrated that the XC7VH290T-2-HCG1155 Xilinx Kintex-7 devices could be used as the fastest FPGA chip device. A detailed comparative analysis also indicates that in terms of area and time the proposed Optimized-GCDSAD method based on SAD block outperforms previously known results.

## I. Introduction

The development of cloud computing and its applications can help to incorporate huge amounts of data generating every second in which they need to be stored, analyzed and secured. The security of this big data against the wide spread of cyber-attacks and serious threats is one of the significant problems for users and service providers. Cryptographic technique is an effective measure that can guarantee secured communication channels for such purposes. The GCD (Greatest Common Divisor) operation is one of the well-known operations that dominates the performance of many public-key crypto-processors [1]. Several algorithms have been applied to find the GCD (Greatest Common Divisor) of two positive numbers. The Euclidean algorithm is one of the well-known solution that has been widely used to compute the GCD. In the last decade, several methods have been proposed to address the design and implementation of an efficient Euclidean algorithm. One of the popular one is the hardware design using FPGA technology [11] with high level simulations and synthesize. While, the instruction set of a conventional 64-bit CPUs cannot support integers of more than 64-bit in length, using hardware device such as FPGA, VLSI or GPU can be an efficient way to implement the arithmetic operations on large integers [11]. Field Programmable Gate Arrays (FPGAs) are semiconductor devices that are based around a matrix of configurable logic blocks (CLBs) connected via programmable interconnects that can implement various digital systems using hardware description languages (HDLs) such as VHDL programming and design synthesize. FPGAs can be reprogrammed to desired application or functionality requirements after manufacturing. Since the continuing decline in the ratio of FPGA price to performance and its programmable features, FPGA is suitable for a hardware implementation of many fields and applications such as high performance computing, networking, security and cryptography, fault tolerance application and many other fields. Computation of the GCD of very large integers is heavily used in computer algebra systems, cryptography, data security and other important algorithms [11]. Our focus in this paper is to propose a new method named Optimized-GCDSAD that computes the GCD of two 32-bit numbers using Sum of Absolute Difference (SAD) block which is based on a fast carry-out generation function. Furthermore, comparative performance evaluation of the FPGA implementation for GCD using five different FPGA chip families, and the synthesize results related to the area of the design, the total delay of the design, minimum delay, maximum frequency are provided. We will demonstrate that the use of new FPGA chip technology offers better hardware utilization and improves performance.

The organization of this paper is described as follows: Section II, discusses the related works on GCD designs and implementations. Notations and preliminaries of Euclidean's GCD algorithm have been provided in Section III. In Section

IV, the architecture of the proposed Optimized-GCD algorithm has been investigated. The architectural complexity and the performance comparison are discussed in Section V. Finally, the conclusion remarks are given in Section VI.

## II. LITERATURE REVIEW

Numerous techniques have been proposed in the literature to implement GCD algorithm efficiently. In [2] four different algorithms, namely Euclid's method, the divisibility check method, the dynamic modulo method and the static method for 8-bit GCD processor have been suggested. They showed that the dynamic modulo algorithm can improve time complexity of GCD processor, whereas Euclid's method was the best method in terms of space complexity. In [3] authors proposed a fast GCD coprocessor based on Euclid's method with variable precisions (32-bits to 1024-bits). They implemented their proposed method on one Altera chip and six Xilinx chips. The experimental results showed that the Xilinx vertex-7 and Xilinx Kintex-7 are the fastest FPGA chip devices in their proposed implementation, which recorded the best maximum frequencies of 243.934 MHz for 32-bit precision and 39.94 MHz for 1024-bit precision. In [4] BIST technology based on Euclid's and Stein's algorithms for 4-bit and 8-bit GCD processors has been proposed. They implemented their proposed method on three Xilinx Spartan 3 devices, including the XC3S50, XC4VFX12 and XC6SLX4. Their experimental results showed that in terms of hardware complexity the XC6SLX4 was the most efficient device. In [5] the authors proposed a ALU based on BIST method using Euclid's and Stein's algorithms to compute 8-bit GCD of two numbers. They implemented their proposed method using different Xilinx families with and without BIST technique and showed that Spartan 3E FPGA family had the lowest power dissipation of 34mW.

In [11] has been proposed FDFM approach for Euclidean algorithm on the FPGA. In their proposed processor core that named GCD processor core, only one DSP slice and one block RAM have been used, and 1280 GCD processors was implemented in a Xilinx Virtex-7 family FPGA XC7VX485T-2. They showed that the performance of this FPGA implementation using 1280 GCD processor cores is 0.0904µs per one GCD computation for two 1024-bit integers, which is 3.8 times faster than the best GPU implementation and 316 times faster than a sequential implementation on the Intel Xeon CPU. The same implementation was applied in [12] with 1408 processors that computed the GCD in parallel. The results showed that GCD computation of two 1024-bit RSA moduli is run at 0.057µs by their proposed new core which is 6 times faster than the best GPU implementation and 500 times faster than a sequential implementation on an Intel Xeon CPU. In this paper, we propose a new Euclidean's GCD algorithm based on Sum of Absolute Difference [7, 9] that is available in the literature. For more area-efficient implementation, the proposed algorithm eliminates the logical XOR gates in carry-out generation block using a well-known logical relation. Due to the fact that NAND gate has lower area and time complexities as compared to other gate complexities such as AND or XOR/XNOR, employing this efficient logical relation can lead to hardware efficiency and lower critical path delay [6]. Analysis shows that the proposed GCD algorithm achieves low-area and low-delay compared to the majority of similar GCD algorithm available in the literature.

## III. NOTATIONS AND PRELIMINARIES

One of the well-known method for computing the GCD (Greatest Common Divisor) of two positive numbers is Euclidean algorithm or Euclid's algorithm [8]. Euclid's algorithm begins with a pair of positive integers and makes a new pair consisting of the smaller number and the difference between the smaller and larger numbers. This process repeats until a zero remainder is obtained. That number then is the greatest common divisor of the original pair. Euclid algorithm can be described as follow [8]:

$$\gcd(a,b) = \begin{cases} a & if \quad a = b \\ \gcd(a-b,b) & if \quad a > b \\ \gcd(a,b-a) & if \quad a < b \end{cases} \quad (1)$$

To obtain the GCD, equation (1) can be applied repetitively to reduce the values of two operands. This equation can also be implemented using for-loop statement using VHDL code.

## IV. PROPOSED LOW-COMPLEXITY OPTIMIZED_GCDSAD ALGORITHM

One of the well-known method for hardware implementation of Euclidean GCD is the GCD2SUB architecture which is based on data-path and control units [3]. The design of the GCD calculator has been divided into two main parts: a controller and a datapath. The controller is an FSM (finite state machine) that issues commands to the datapath based on the current state and the external inputs. This can be described using behavioral model in VHDL. The datapath design is based on structural model containing a netlist of functional units or modules such as multiplexers, registers, subtractors, and a comparator.

The controller steps through the GCD2SUB algorithm has been shown in Figure 1. If $x = y$, the computing of GCD would be finished. It can be seen that the block diagram of the GCD2SUB contains a control unit and a data path unit. Data path unit consists of two 2-to-1 multiplexers, three registers, two subtractors and one comparator. Each of the components is compiled and simulated separately. Furthermore, a control unit is implemented using behavioral technique and FSM using VHDL code. Finally, the GCD calculator can be implemented in top-level design using structural model in which port map instruction interconnects the data path and control unit.

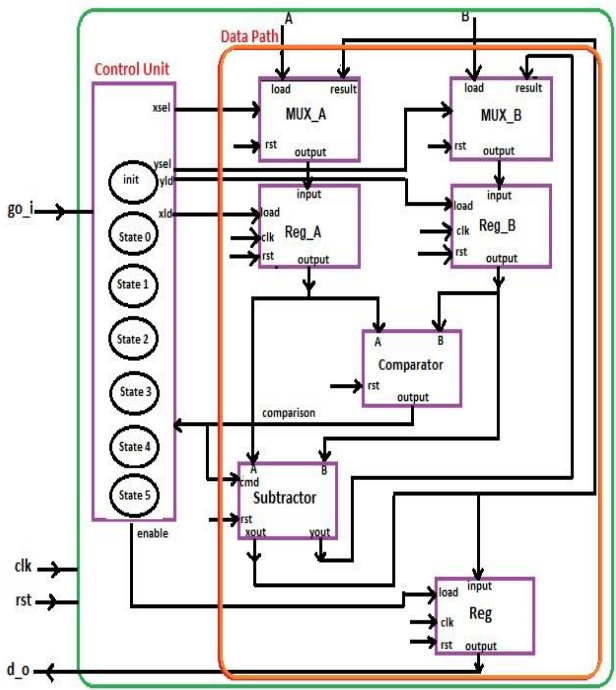

Figure 1: The block diagram of GCD2SUB method

To area-efficient implementation of Eculidean's GCD algorithm, subtractors and comparators in GCD2SUB architecture can be replaced by Sum of Absolute Different (SAD) using the carry-out generation function. The architecture of the proposed GCD that we called GCDSAD has been illustrated in Figure 2 consisting of two 2-to-1 multiplexers, three registers, one SAD block and one FSM.

The general algorithm for computing the Sum of Absolute Difference (SAD) of two numbers consists of the following steps [7, 9]:

**Step1: Determining the smallest of the two operands**: As suggested in [9], it is only necessary to determine whether $A + \bar{B}$ produces a carry or not. This can be done using the carry-out Generation Function. There are two cases resulting from this function:

**I: No carry was generated**: This implies $B \prec A$. In this case the inversion of $B$ to $\bar{B}$ should be applied. The value of $B$ is equal to the positive number $2n - 1 - B$. This number is again in unsigned binary representation. The value $A$ should be propagated unmodified. Their sum equals $2^n - 1 - B + A = 2^n - 1 + |A - B|$.

**II: carry was generated**: This implies $B > A$. In this case, we should invert $A$ to $\bar{A}$ and propagate $B$ unmodified. Their sum equals $2^n - 1 - A + B = 2^n - 1 + |A - B|$.

**Step 2: Inverting the smallest value**: If no carry was produced, $B$ must be inverted, otherwise, $A$ must be inverted. This is done using one XOR gate. Moreover, to check the equivalence of two numbers, a 32-bit NOR gate has been adopted in the output of the SAD circuit.

In Figure 3, the block diagram of the carry generation is depicted. The determination of whether the addition $A + \bar{B}$ generates a carry, is performed without actually calculating the addition. Instead, this is achieved by only utilizing certain parts within a carry look-ahead adder that calculates the carry. The resulting carry and inverted carry are fed to one XOR that will invert the correct term.

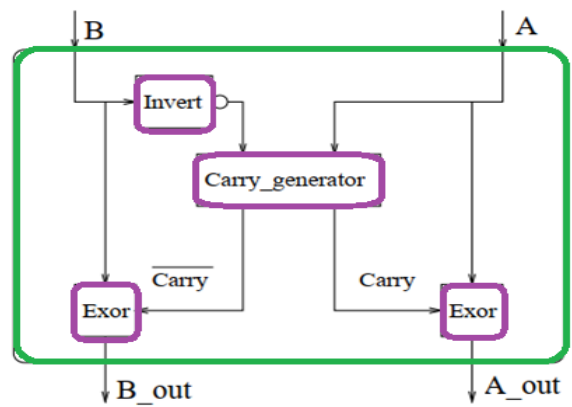

Figure 3: The block diagram of the carry generation

As it is mentioned, the carry-out generation function with XOR gate has been used in the SAD (Sum of Absolute Difference) block. For more efficient implementation, all logical operations in the SAD block can be converted to an equivalent circuit that only uses NAND in a particular architecture that we called Optimized-GCDSAD, it can decrease the cost of SAD and increase the speed of the SAD

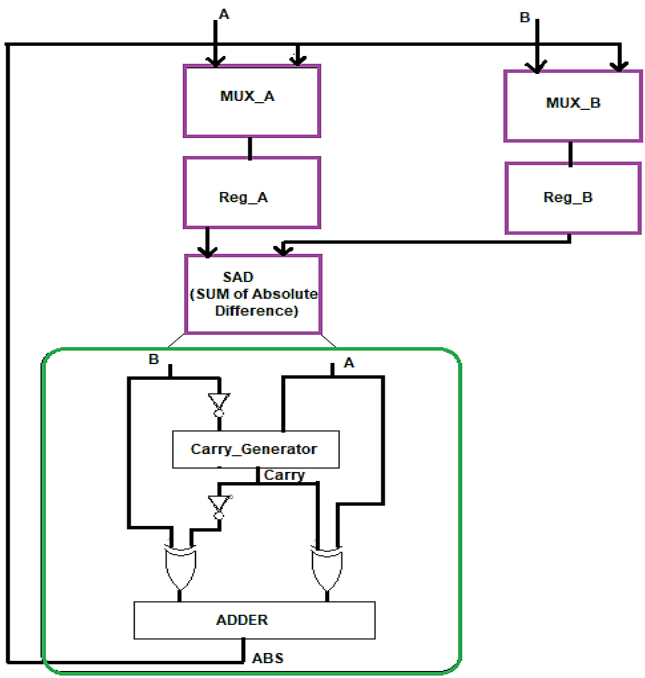

Figure 2: The block diagram of GCDSAD method

block [10]. For any two logic variables *a* and *b*, we can find that:

$$NOT\ a = a\ NAND\ b$$
$$a\ AND\ b = \overline{(a\ NAND\ b)\ NAND\ (a\ NAND\ b)} \quad (2)$$
$$a\ OR\ b = \overline{(a\ NAND\ a)\ NAND\ (b\ NAND\ b)}$$
$$a\ XOR\ b = \overline{(a\ NAND\ (a\ NAND\ b))\ NAND\ ((a\ NAND\ b)NAND\ b)}$$

From equation (2), the carry outputs of each stage can be expressed as follow:

$$NB = \overline{BB}$$
$$P = \overline{\overline{NB.NB.A.A}.\overline{NB.A}}$$
$$G = \overline{NB.A..NB.A} \quad (3)$$
$$P4(i) = \overline{P(4i-1).P(4i-2).P(4i-3).P(4i-4)}$$
$$G4(i) = \overline{G(4i-1).\overline{(G(4i-2).P(4i-1))}.\overline{(G(4i-3).P(4i-1).P(4i-2))}.}$$
$$\overline{G(4i-3).P(4i-1).P(4i-2).p(4i-3)}$$
$$C = \overline{G4(4).\overline{(G4(3).P4(4))}.\overline{(G4(2).P4(4).P4(3))}.}$$
$$\overline{(G4(1).P4(4).P4(3).P4(2))}$$

Furthermore, according to Figure 3, the resulting carry and inverted carry which are fed to the XOR gate for inverting the correct term, can be expressed as below:

$$BX(i) = \overline{\overline{B(i).(\overline{B(i).C}).(\overline{B(i).C}).C}} \quad (4)$$
$$AX(i) = \overline{A(i).(\overline{A(i).C}).(\overline{A(i).C}).C.\overline{A(i).(\overline{A(i).C}).(\overline{A(i).C}).C}}$$

These equations reveal that to compute the desired GCD result, the logical NAND operation can be used instead of XOR, AND and OR logical operations. Compared to the conventional equation, this method can significantly improve hardware and time complexity that will be explain in the results section.

## V. FPGA IMPLEMENTATION RESULTS

The architecture of the proposed algorithm is modelled using VHDL for numbers of 32-bit size, and then simulated and synthesized using Xilinx ISE 14.7 to verify the functionality. The proposed 32-bit Optimized-GCDSAD algorithm was synthesized using five different FPGA chips such as the Xilinx Virtex-7 (XC7VH290T-2-HCG1155), the Xilinx Virtex-5(XC5VlX20T-2-FF323), the Xilinx Spartan-6 (XC7Z010-2-CLG400), the Xilinx Kintex-7 (XC7K70T-2-FBG676), and the Xilinx Zynq (XC7Z010-2-CLG400). The bar chart in Figure 4 compares the maximum frequency values (in MHz) (x-axis) of implementation lengths 32 bit for the five FPGA devices (y-axis) of our proposed method and the proposed GCD processor in [3]. At the first glance it is evident that the maximum frequency value of our proposed optimized-GCDSAD is more than the proposed method in [3] in almost all FPGA chips. While the highest frequency rate of our method belongs to Virtex-7 and Virtex-5 with 303.893 MHz which are %89.93 and %28.56 higher than their counterparts in [3] respectively, the lowest frequency rate lies to Spartan-3 with 128.131 MHz that is %0.05 higher than [3]. The Kintex-7, and the Xilinx Zynq occupy the second place frequency rate with 271.467 MHz which are %11.3 and %20 greater than their counterparts in [3], respectively.

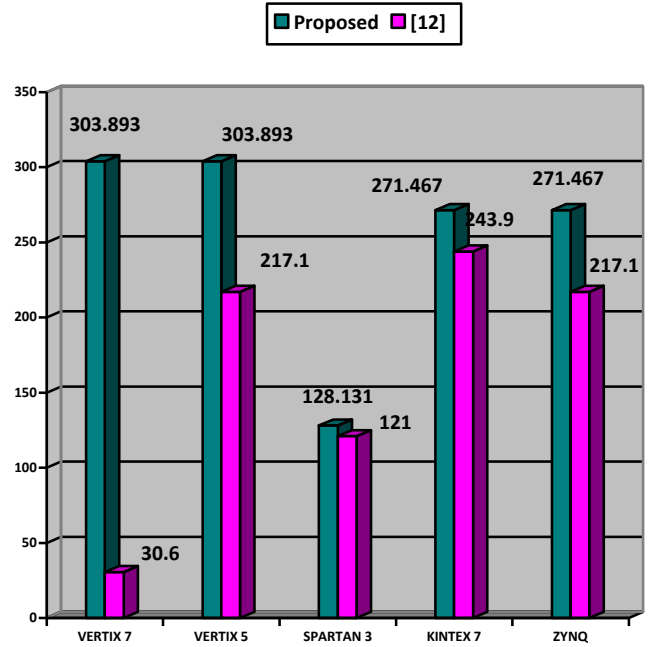

Figure 4: Maximum Frequency Values (MHz)

Figure 5 shows the minimum period value (x-axis) of our method and the proposed method in [3] for five FPGA devices (y-axis). It is obvious that the minimum period value of our method is lower than that of in [3], in all device families. Another interesting point is that, in our proposed method, the lowest minimum period values belong to Vertix-7 and Kintex-7 with 3.2 ns. So, the proposed GCD-optimized method with Vertix-7 and Kintex-7 device family can be a good candidate for implementation of the proposed 32-bit GCD.

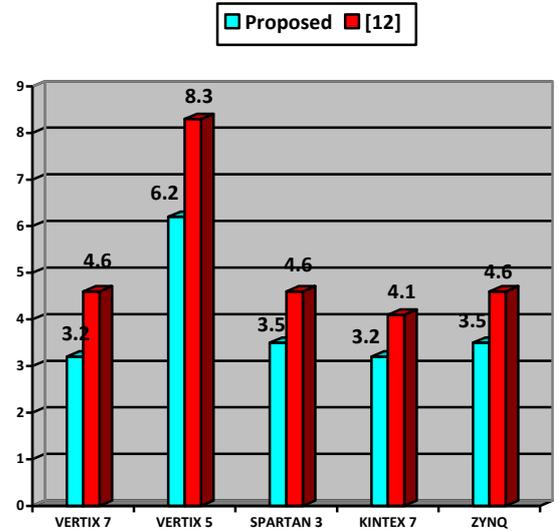

Figure 5: Minimum Period Values (*ns*)

Figure 6 illustrates the maximum delay value (x-axis) of our method and the proposed method in [3] for five FPGA devices (y-axis). It is clear that the maximum delay value of our method is lower than that of in [3], in all device families. It should also be noted that, in both proposed methods, the best delay time belongs to Kintex-7 with 4.8 ns and 8.2 ns in our method and in [3], respectively. Therefore, our GCD processor based on SAD block with the same chip family Kintex-7 computes the GCD approximately 2 times faster.

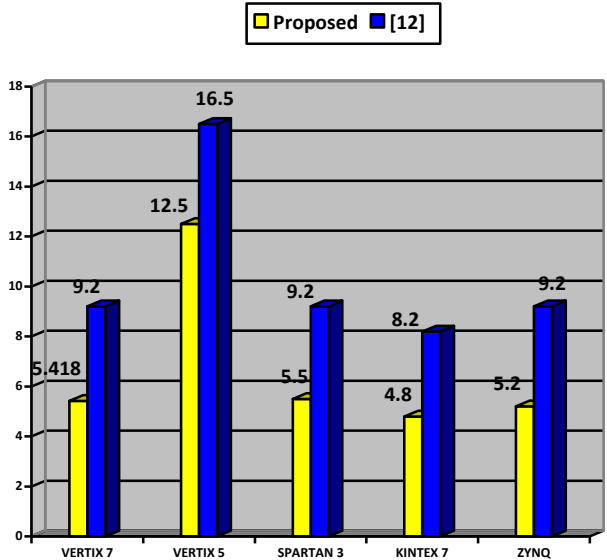

Figure 6: Maximum Delay Values (*ns*)

Furthermore, the synthesized netlist is implemented on FPGA Xilinx Vertix-7 (XC7VH290T-2-HCG1155) prototype board. The delay, area, ADP (area-delay product) obtained from the synthesis results are tabulated in Table 1. Moreover, the area and time complexities of the proposed 32-bit Optimized-GCD algorithm are compared with other three architectures including GCDSAD, GCD2SUB, and GCD for-loop. Although the overall structure of the GCDSAD and GCD2SUB architectures might seem similar, there are two important differences between them which is, the implementation of the proposed GCDSAD with Sum of Absolute Difference which is based on a fast carry-generation function that can be more area-efficient than GCD2SUB design. From the results shown in Table1, GCD-FOR-LOOP has more number of components, whereas, the proposed Optimized-GCDSAD requires the least number of LUTs and registers resulting in less area requirement.

In the design of digital circuit, there is always a trade-off between delay and area costs as two important design factors and reducing one of them generally results in an increase in the other one. The area-delay product (ADP) parameter is the balanced parameter for overall performance comparison of various architectures rather than the area and delay individually. From the results shown in Table 2, GCD_FOR _LOOP architecture has long latency among all. It can also be observed from Table 2 that GCD2SUB takes less delay at the expense of more number of LUTs and registers compared to the GCDSAD and Optimized-GCDSAD. The proposed Optimized-GCDSAD algorithm achieves the best area-delay product compared to other three methods considered for comparison. It is evident from Table 2 (% reduction in ADP row) that the proposed Optimized-GCDSAD achieves area-delay efficiency of 15%, 6% and 81% when compared with GCDSAD, GCD2SUB, and GCD-FOR-LOOP, respectively. Hence, it is clear from the estimation values presented in Table 2 that the Optimized-GCDSAD algorithm is area and area-delay product efficient.

**Table 1**. Area complexities comparison between four Euclidean's GCD architectures using Xilinx Kintex-7 (XC7VH290T-2-HCG1155) device

| Hardware Factor | Optimized-GCDSAD Utilization | GCDSAD Utilization | GCD2SUB Utilization | GCD-For-loop Utilization |
|---|---|---|---|---|
| #Slice Registers | 107/ 184304 (0%) | 107/ 184304 (0%) | 107/ 184304 (0%) | 0/ 184304 (0%) |
| #Slice LUTs | 156/ 92152 (0%) | 216/ 92152 (0%) | 301/ 92152 (0%) | 19205/ 92152 (20%) |
| #LUT_FF pairs | 75/ 303 (30%) | 75/ 303 (30%) | 105/ 303 (34%) | 0/ 303 (0%) |
| #bonded IOBs | 99/ 396 (25%) | 99/ 396 (25%) | 99/ 396 (25%) | 96/ 396 (32%) |
| #BUFG/ BUFGCTRLs | 1/16 (6%) | 1/16 (6%) | 1/16 (6%) | 1/16 (6%) |

**Table 2**. Delay Time and ADP comparison between four Euclidean's GCD architectures

| Macro Statistics | Optimized-GCDSAD | GCDSAD | GCD2SUB | GCD-FOR-LOOP |
|---|---|---|---|---|
| Total Delay Time | 6.579 ns | 6.311 ns | 4.529 ns | 492.707 ns |
| Area * Delay (ADP) | 1727 | 2038.453 | 1847.832 | 9462437.935 |
| %Reduction in ADP | - | %15 | %6 | %81 |

Table 3 presents the comparison of estimated area, delay, area-delay product (ADP) of the proposed Optimized-GCD with the proposed GCD in [3]. It is observed that the proposed Optimized-GCD involves nearly 60.04% less area-delay product (ADP) and is about 1.7 ns faster with compared to proposed implementation of 32-bit GCD in [3]. Hence, the FPGA implementation results obtained from the proposed 32-bit GCD validate hardware overhead, time complexity, and area-delay product (ADP) complexity advantage over the proposed method in [3].

**Table 3**. Comparison of FPGA implementation between the proposed method and [3]

| Design | Area (#LUTs) | Delay (*ns*) | ADP (#LUT × Delay) | %Reduction in ADP |
|---|---|---|---|---|
| Proposed method | 256 | 6.5 *ns* | 1664 | - |
| [3] | 508 | 8.2 *ns* | 4165.6 | %60.04 |

## VI. CONCLUSION

In this paper proposed an efficient FPGA implementation of 32-bit GCD Euclidian's algorithm using Sum of Absolute Difference named Optimized-GCDSAD. It should also be noted that the complexity of this method is achieved by means of utilizing logic NAND gate in a fast carry-out generation function in a particular architecture. The proposed method was implemented and synthesized using five different Xilinx chips. The conducted experiments showed that Xilinx Kintex-7 (XC7VH290T-2-HCG1155) is best-suited FPGA chip device in terms of speed. Furthermore, our proposed Optimized-GCD offers 60.04%, area-delay improvement (ADP) and is 1.7 ns faster than the proposed GCD in [3], and it can be embedded with many FPGA based cryptographic applications.

REFERENCES

[1] A. Mohammed Musab, M. Ibrahim, A. Al-Haija, Q. Alshuaibi. Abdullah, "Performance Analysis of 128-bit Modular Inverse Based Extended Euclidean Using Altera FPGA Kit", Procedia Computer Science. 160. 543-548. 10.1016/j.procs.2019.11.050, 2019.

[2] D. Upadhyay, J. Kolte, K. Jalan, "Approach to design Greatest Common Divisor Circuits based on Methodological analysis and Valuate Most Efficient Computational Circuit", International Journal of Electrical and Electronics Engineering Research (IJEEER), Vol 3(4); Pp. 59-66, 2013.

[3] Q. Abu Al-Haija, S. A. Sharifah Mumtazah, I. Alfarran, "FPGA Implementation of Variable Precision Euclid's GCD Algorithm", Journal of Engineering Technology. 6. 410-422, 2017.

[4] S. D. Kohale, R. W. Jasutkar, "FPGA Based Implementation of BIST Controller Using Different Approaches", International Journal of Materials, Mechanics and Manufacturing, vol. 1 (2); Pp 110-113, 2013.

[5] S. D. Kohale, R. W. Jasutkar, "Power Dissipation of ALU Implementation of GCD Processor with and Without BIST Among Various Xilinx Families", International Journal of Engineering Research & Technology (IJERT), Vol. 2(2); Pp 1-7, 2013.

[6] S. Nabipour, Gh. Zare Fatin, J. Javidan, "Area- Efficient VLSI Implementation of Serial-In Parallel-Out Multiplier Using Polynomial Representation in Finite Field GF($2^m$)", arXiv preprint arXiv:2007.08284, https://arxiv.org/abs/2007.08284, 2020.

[7] S. Vassiliadis, E. Hakkennes, S. Wong, and G. Pechanek, "The Sum-Absolute-Difference Motion Estimation Accelerator", In *Proceedings of the 24th Euromicro Conference*, 2000.

[8] W. JSSM., Vassiliadis, S., & Cotofana, SD, "A sum of absolute differences implementation in FPGA hardware" In M. Fernandez (Ed.), *EUROMICRO 2002;* Proceedings of the 28th EUROMICRO Conference (pp. 183-188). IEEE Society, 2002.

[9] W. Trappe, L. C. Washington, "Introduction to Cryptography with Coding Theory", Prentice Hall, Vol (1); Pp. 1-176, 2002.

[10] Y. Pai, Y. Chen, "The fastest carry lookahead adder" Proceedings. DELTA 2004. Second IEEE International Workshop on Electronic Design, Test and Applications, Perth, WA, Australia, pp. 434-436, doi: 10.1109/DELTA.2004.10071, 2004.

[11] Z. Zhou, K. Nakano, Y. Ito, "Efficient Implementation of FDFM Approach for Euclidean Algorithms on the FPGA", International Journal of Networking and Computing, Vol 6 (2); P.p. 420–435, 2016.

[12] Z. Zhou, K. Nakano, Y. Ito, "Parallel FDFM Approach for Computing GCDs using the FPGA", Springer: PPAM 2015, Part I, LNCS 9573, pp. 238–247, DOI: 10.1007/978-3- 319-32149-3 23, 2016